\documentclass[pra,aps,graphics,a4paper,10pt,twocolumn]{revtex4-1}

\usepackage[margin=15mm]{geometry}

\usepackage{float}
\usepackage{graphicx}
\usepackage[caption=false]{subfig}
\usepackage{mathrsfs}
\usepackage{xfrac}
\usepackage{enumitem}
\usepackage{epstopdf}
\usepackage{amsmath}
\usepackage{amssymb}
\usepackage{braket}
\usepackage{color}
\usepackage{fp}
\usepackage{placeins}
\usepackage{array}
\usepackage{inputenc}
\usepackage{natbib}
\usepackage{hyperref}
\usepackage{tabularx}
\usepackage{tabu}
\usepackage{colortbl}
\usepackage[dvipsnames]{xcolor}
\usepackage{soul}
\usepackage{algorithmic}
\usepackage{ulem}
\usepackage{multirow}
\usepackage{bbold}
\usepackage{mathtools}

\newcommand{\be}{ \begin{equation}}
\newcommand{\ee}{ \end{equation}  }

\begin{document}
	\title{Engineered Nearest-Neighbour Interactions with Doubly Modulated Optical Lattices}

	\author{Hongzheng Zhao}
	\author{Johannes Knolle}
	\author{Florian Mintert}
	\affiliation{Physics Department, Blackett Laboratory, Imperial College London, Prince Consort Road, SW7 2AZ, United Kingdom}
	
	\begin{abstract}
		Optical lattice systems provide exceptional platforms for quantum simulation of many-body systems. We focus on the doubly modulated Bose-Hubbard model driven by both time-dependent on-site energy and interaction, and predict the emergence of the nearest neighbour interaction and density-assisted tunnelling. By specifically designing a bi-chromatic driving pattern for a one dimensional lattice, we demonstrate that the doubly modulated fields can be tuned to realize desired quantum phases, e.g. the Mott insulator phase with selective defects, and density wave phase.
	\end{abstract}  
	
	\maketitle
	\section{Introduction}
	
The rapid progress in accurate control of matter by means of laser fields
enables the realization of quantum simulations~\cite{QS} in various controllable systems, such as optical lattices with atomic~\cite{OLBHM} or molecular gases~\cite{Dipolar}, and trapped ions~\cite{RTLGT}.
One of the central tools for the realization of desired Hamiltonians is Floquet engineering~\cite{HF,SD}
that gives access to the simulation of physical processes that would be prohibitively difficult to realize by static means~\cite{NS,Mag1}.
With the help of rapid periodic shaking of optical lattices it is possible to tune the amplitudes of quantum tunnelling between nearest neighbours ~\cite{DC} and beyond that, to create long-range tunnelling processes~\cite{SD} and modify topological properties~\cite{TC}.
Periodically modulated external magnetic fields can be employed in the vicinity of Feshbach resonances~\cite{FR3,FR4} for the dynamical control of on-site interactions in optical lattices.
The tunability of such interactions enabled the observation of the phase transition from superfluid to Mott insulator \cite{SIT}, and it can also be exploited for the realisation of density assisted tunnelling processes~\cite{EU,CH}.
	
	So far, most activities in driven optical lattices are concerned with the control of tunnelling properties or on-site interactions. 
	Only recently, also control of
	nearest neighbour interactions with great importance for the realization of entangled many-body ground states~\cite{EBHM,QP,Dipolar,LR1,Entangle} or far-from-equilibrium quantum many-body dynamics~\cite{cor,LR2} came into the spotlight.
In principle, effective interactions beyond on-site interactions can be realised in terms of regular lattice shaking, but those effective interactions will typically be much smaller than the intrinsic on-site interactions.
In order to overcome this, we consider, in addition to a periodic shaking, also a periodic modulation of the on-site interaction.
We derive a complete classification of effective processes in this system and find driving profiles that allow us to selectively enhance or suppress such processes.
With the explicit example of a density-wave we demonstrate what type of physics~\cite{BHM1,BHMSFMI, CDW1, HI,EBHM} can be simulated in terms of such a doubly-modulated Bose-Hubbard system.

	\section{The Doubly Modulated Model}
	\label{Sec.DM}
	A system of spinless bosons occupying the lowest band in an optical lattice can be described by the Bose Hubbard Hamiltonian
		\begin{equation} 
	\label{equ.H1}
	\hat{H}(t) = \sum_{\langle jk\rangle}\hat{c}_j^{\dagger}J_{jk}\hat{c}_k+\frac{U(t)}{2}\sum_i \hat{n}_j(\hat{n}_j-1)+F(t)\sum_{j}\epsilon_j\hat{n}_j
	\end{equation}
	where $\hat{c}_j(\hat{c}_j)^{\dagger}$ annihilates (creates) a boson at site $j$, $\hat{n}_j=\hat{c}_j^{\dagger}\hat{c}_j$ is the occupation number operator.
	$U(t)$ is the amplitude of the on-site interaction,
	$F(t)$ is the amplitude of tilt or shaking, $\epsilon_j$ is given by the lattice geometry
	and $J_{jk}$ is the bare tunnelling rate between nearest neighbours, {\it i.e.} the summation $\sum_{\langle jk\rangle}$ is performed over adjacent sites only. 
	We will consider the doubly modulated version of the Bose-Hubbard Model(BHM)~\cite{EU}, in which both on-site energy and the on-site interactions have a periodic time-dependence with period $T$.
	In practice, these time-dependences result from a periodic tilt or acceleration of the lattice, and 
	a modulation of a magnetic field inducing a Feshbach resonance~\cite{FR4,FR3}.
	We will consider a vanishing average on-site energy, {\it i.e.} $\int_{t}^{t+T}F(\tau)d\tau=0$, but allow for a finite average interaction $U_0=\int_{t}^{t+T}U(\tau)d\tau/T$, and denote the periodic deviation from the average by $U_f(t)$.

For a periodically driven system, Floquet theory~\cite{Flo1,Flo2,Flo3} permits to describe the dynamics in terms of an effective time-independent Hamiltonian $\hat{H}_{eff}$ that induces the propagation over one period of driving.
	If the driving frequency $\omega$ exceeds all system energy scales the effective Hamiltonian $\hat{H}_{eff}$ can be computed in a perturbative manner \cite{Mag1,HF} as sketched in Appendix~\ref{Sec.effectiveH}.
	Since strong driving is conflicting with this condition, it is necessary to consider the system Hamiltonian in a frame in which the strong driving amplitude does not contribute to the magnitude of the Hamiltonian, {\it i.e.} the energy scales of the system.
	It is necessary for the shaking $F(t)$ to be strong ({\it i.e.} $F(t)\gtrsim \omega$), but the modulation of the interaction does not need to exceed the driving frequency.
	A possible choice for the unitary defining a suitable frame would thus be 
	$\exp(i\sum_j\theta_j(t)n_j)$ with $\theta_j(t) = \epsilon_j\int_{0}^{t}d\tau\ F(\tau)$ as frequently employed in regular shaken lattices~\cite{HF}.
	The unitary
	\begin{equation}
	\label{equ.unitary}
	U_I(t) =\exp\left(i\left(\sum_j\theta_j(t)n_j +\frac{\Gamma(t)}{2}\sum_j  n_j(n_j-1)\right)\right)\ ,
	\end{equation}
	with $\Gamma(t)=\int_{0}^{t}d\tau\ U_f(\tau)$
	will however ease the analysis and result in more compact expressions.
	Strictly speaking, the frame defined by $\hat{U}_I(t)$ would enable a treatment even if the amplitude of the time-dependent interaction exceeds the driving frequency $\omega$.
	 Since this, however would result in new effective many-body interactions, we will limit our analysis to sufficiently weak interactions, and apply a corresponding series expansion where appropriate.

	The transformed Hamiltonian
	$\tilde{H}(t)=U_I\hat{H}(t)U^{\dagger}_I-iU_I \dot U^{\dagger}_I$
		in the frame defined by $U_I$ reads
	\begin{eqnarray}
	\label{equ.H}
	\begin{aligned}
	\tilde{H}(t)=\sum_{\langle jk\rangle} \hat{c}_j^\dagger \hat{A}_{jk}(t) \hat{c}_k+\frac{U_0}{2}\sum_{j}\hat{n}_j(\hat{n}_j-1)\ ,
	\end{aligned}
	\end{eqnarray}
	and admits a perturbative expansion even if the amplitude $F(t)$ exceeds the driving frequency $\omega=2\pi/T$.
	The system's rich physics is contained in the time-dependent operators~\cite{TC,SIT}
	\begin{eqnarray}
	\label{eq.dynamicalA}
	\hat{A}_{jk}(t) =J_{jk}e^{i\chi_{jk}(t)}e^{i\Gamma(t)(\hat{n}_j-\hat{n}_k)}\ ,
	\end{eqnarray}
	defined in terms of the displacements $\textbf{r}_{jk}=\textbf{r}_j-\textbf{r}_k$, and the phase $\chi_{{jk}}(t)=\theta_k(t)-\theta_j(t)$, that a particle acquires when tunnelling from site $k$ to site $j$.

At the first glance, $\tilde{H}(t)$ given in Eq.~\eqref{equ.H} looks like a standard Bose-Hubbard Hamiltonian.
The first term of $\tilde{H}(t)$, however, does not describe only single-particle tunnelling from site $k$ to site $j$,
but the term $\hat{A}_{jk}^{0}$ may be interpreted as density assisted tunnelling, {\it i.e.} the amplitude for a tunnelling process between sites $j$ and $k$ depends explicitly on the particle number difference between those two sites.
The physics described by $\tilde{H}(t)$ is thus substantially richer than that of the bare Bose-Hubbard Hamiltonian. 

	\section{Effective Processes}
	\label{Sec.Eff}
	
	Since the perturbative construction of the effective Hamiltonian is entirely analytic, it can be done without specifying an explicit time-dependence of the system.
	We will therefore leave the driving profiles $\Gamma(t)$ and $\theta_j(t)$ unspecified and determine specific time-dependencies only later in Sec. \ref{Sec.D}.
	
Standard perturbative expansions provide the effective Hamiltonian as series in powers of $J_{jk}/\omega$ and $U_0/\omega$,
and an expansion including terms up to $1/\omega^2$ is appropriate for our purposes.
Within this approximation, the effective Hamiltonian still contains general dependence on the amplitudes of shaking and the modulation of on-site interactions.
Since the latter needs to be assumed to be sufficiently small, we can perform the corresponding series expansion and neglect terms beyond order $1/\omega^2$.
Wherever, such an expansion is not helpful, we will keep the general dependence, but bear in mind that sufficiently weak interactions are assumed.

The lowest order ($1/\omega^0$) effective Hamiltonian is a regular Bose-Hubbard Hamiltonian with modified tunnelling and interaction rate~\cite{HF,SIT,Colloquium}.
In first order, there is density assisted tunnelling (in terms of the operator $c_j^\dagger(\hat n_j-\hat n_k)c_k$) as a physically distinct new process.
Of particular interest for our present purposes is the second order ($1/\omega^2$) nearest neighbour interaction $\hat n_j\hat n_k$ resulting from a particle tunnelling from site $j$ to site $k$, interacting with the particles at site $k$ and tunnelling back to site $j$,
as reflected by the commutation relation
\begin{equation}
\label{eq.longrange}
\left[c_j^\dagger c_k,\left[\hat n_k(\hat n_k-1),c_k^\dagger c_j\right]\right]=4\hat n_j\hat n_k-2\hat n_j^2 .
\end{equation}

As discussed in more detail in Appendix.~\ref{Sec.effectiveH} the contributions to the effective Hamiltonian that are of purely interacting nature, result in
	\begin{eqnarray}
	\label{eq.int}
	\begin{aligned}
	\hat{H}^{int}_{eff}=\sum_{\langle jk\rangle}\frac{U_{jk}^{N}}{2}\hat{n}_j\hat{n}_k+\sum_j \frac{U_j^{O}}{2} \hat{n}_j(\hat{n}_j-1)\ ,
	\end{aligned}
	\end{eqnarray}
	where the summation $\sum_{\langle jk\rangle}$ is performed over neighbouring sites only with the interaction constants
	\begin{eqnarray}
	\label{equ.U}
	\begin{aligned}
	U_{jk}^N &= -4\sum_{n\neq 0}\left[\frac{1}{\omega n}(g_{jk}^{-n}t_{kj}^{n}+t_{jk}^{n}g_{kj}^{-n})-\frac{U_0}{2n^2\omega^2}g_{jk}^ng_{kj}^{-n}\right]\ ,\\
	U_{j}^O&={U_0}-\sum_{k}U_{jk}^N\ ,
	\end{aligned}
	\end{eqnarray}
	defined in terms of the Fourier components
	\begin{eqnarray}
	\label{equ.co}
	\begin{aligned}
	g_{jk}^n&=\frac{1}{T}\int_0^T dt\ J_{jk}e^{i\chi_{{jk}}(t)}e^{-in\omega t}\ ,\\
	t_{jk}^n&=\frac{i}{T}\int_0^Tdt\ J_{jk}e^{i\chi_{{jk}}(t)}\Gamma(t)e^{-in\omega t}\ ,
	\end{aligned}
	\end{eqnarray}
 that depend on the driving functions $\Gamma(t)$ and $\theta_j(t)$ (via $\chi_{jk}(t)=\theta_k(t)-\theta_j(t)$).

In addition to these interaction terms and processes contained in the bare system,
the effective Hamiltonian also contains additional tunnelling processes.
The exact form of these processes and their amplitudes depend on the driving profiles similarly to the dependence of the interaction terms given in Eqs.\eqref{equ.U} and \eqref{equ.co}.
The general case is discussed in more detail in Appendix \ref{Sec.effectiveH}.

	\label{Sec.D}
	
In the following, we will exploit the freedom to chose driving profiles in order to realise an effective system with desired properties in a one-dimensional chain.
The profiles can be chosen from a continuous set, but we will see that a monochromatically modulated on-site energy
\be
U_f(t) = 2U_d\cos(2\omega t)
\label{eq:parameterU}
\ee
and a bi-chromatic shaking resulting in 
\be
F(t)=F_1\cos(\omega t)+2F_2\cos(2\omega t)
\label{eq:parameterF}\ee
with $\epsilon_j = -j$ is all that is required.
In order to simplify notation, we will also use the dimensionless quantities $\tilde U_d=U_d/\omega$,  $\tilde F_1=F_1/\omega$ and $\tilde F_2=F_2/\omega$ in the following.


\subsection{Elementary processes}
\label{Sec.process}
Before discussing means to control the system dynamics in terms of the driving parameters $U_d$, $F_1$ and $F_2$ we will first give an overview over all processes found in the effective Hamiltonian up to second order.
Since we employ bi-chromatic driving, the parameters in the effective Hamiltonian can no longer be expressed in terms of regular Bessel functions $J_n(x)$ of a single variable, but we will encounter two-dimensional Bessel functions;
those can be expressed in terms of one-dimensional Bessel functions via the relation
		$J_n(x,y) = \sum_{s=-\infty}^{+\infty} J_{n-2s}(x)J_s(y)$,
		and for all practical purposes the summation can be limited to a finite range.
	
\subsubsection{Nearest-neighbour density assisted tunnelling}
\label{sec.nndat}
The Hamiltonian
\be
\hat{H}_t = \sum_{jk}c_j^\dagger\hat{A}^0_{jk}c_k, 
\ee
given as the temporal average of the tunnelling term in Eq.\ref{eq.dynamicalA},
depends on the driving parameters via the relation
	\begin{equation}
	\label{eq.A}
	\hat{A}^0_{jk} =
	-j_0J_0(\tilde F_1(j-k),\tilde F_2(j-k)+\tilde U_d(\hat{n}_j-\hat{n}_k))\ .
	\end{equation} 
	Because of the term $\hat n_j-\hat n_k$, the prefactors $\hat{A}^0_{jk}$ should not be taken literally as a rate but a proper rate can be specified for given initial and final states only~\cite{EU}.	
	In a one dimensional system with the initial Fock state $\ket{\psi_i}=\ket{n_1\hdots n_N}$ one obtains
	\be
	c_j^{\dagger}\hat{A}^0_{j,j+1}c_{j+1}\ket{\psi_i}=h_{n_j,n_{j+1}}^Lc_j^{\dagger}c_{j+1}\ket{\psi_i}\ ,
	\ee
	with the prefactor
	\begin{equation}
	h^L_{n_j,n_{j+1}}=- j_0J_0(\tilde F_1,\tilde F_2+\tilde U_d(-n_j+n_{j+1}-1))\ ,
	\end{equation} 
	that can be interpreted as the rate for a particle to tunnel from site $j+1$ to $j$, {\it i.e.} to the left.
	The equivalent rate for a particle to tunnel from site $j$ to $j+1$ reads
	\begin{equation}
	h^R_{n_j,n_{j+1}}=- j_0J_0(\tilde F_1,\tilde F_2+\tilde U_d(-n_j+n_{j+1}+1))\ .
	\end{equation}
	States with low occupations per site will be particularly relevant later-on, and in those cases, the above rates reduce to the simple expressions
	\begin{eqnarray}
	\label{equ.channel}
	\begin{aligned}
	h^L_{(0,1)}=h^R_{(1,0)} & =-j_0J_0(\tilde F_1,\tilde F_2)\ ,\\
	h_{(1,1)}^L = h_{(2,0)}^R&=-j_0 J_0(\tilde F_1,\tilde F_2-\tilde U_d)\ ,\\
	h_{(1,1)}^R= h_{(0,2)}^L &=-j_0J_0(\tilde F_1,\tilde F_2+\tilde U_d)\ .
	\end{aligned}
	\end{eqnarray}
	In regular tunnelling processes described by the bare Bose-Hubbard Hamiltonian, the rates to tunnel left or right can only differ by a phase factor,
	but if the factors $h_{j,k}^{L/R}$ are interpreted as a rate, then one needs to accept that rates to tunnel left or right can also differ in their magnitude, as it generally is the case for $h_{(2,0)}^R$ and $h_{(0,2)}$\cite{EU}.
	
	In order to highlight the explicit dependence of particle densities on the tunnelling processes, it is appropriate to expand $\hat{A}^0_{jk}$ (Eq.~\ref{eq.A}) in powers of $U_d$.
The expansion
	\begin{equation}
	\label{eq.A_expand}
	\begin{aligned}
	\hat{A}^0_{j,j+1}
	=- j_0\sum_{\nu=0}^{\infty} \frac{\tilde{U}^\nu_d(\hat{n}_j-\hat{n}_{j+1})^\nu}{\nu !} \frac{\partial^\nu }{\partial \tilde F_2^\nu}J_0(\tilde F_1,\tilde F_2),
	\end{aligned}
	\end{equation}
then yields explicit rates in terms of derivatives of the scalar Bessel function $J_0(\tilde F_1,\tilde F_2)$.
The lowest order term describes regular tunnelling, and the higher order terms correspond to density assisted tunnelling processes.

	\subsubsection{Nearest-neighbour interactions}	
	\label{sec.int}
	In the one dimensional chain the interaction terms of the effective Hamiltonian given in Eq.~\eqref{eq.int} reduces to
	\begin{eqnarray}
	\label{eq.intfor1d}
	\begin{aligned}
	\hat{H}^{int}_{eff}=V_e\sum_{ j}\hat{n}_j\hat{n}_{j+1}+\frac{U_e}{2} \sum_j \hat{n}_j(\hat{n}_j-1)\ ,
	\end{aligned}
	\end{eqnarray}
	with the nearest neighbour interaction constant

		\begin{eqnarray}
		\label{eq.V}
		\begin{aligned}
		V_e=\frac{4j_0^2}{\omega^2}\sum_{n\neq 0}\left[-\frac{U_d}{n}\frac{\partial}{\partial \tilde F_2}+\frac{U_0}{2 n^2}\right]J_{n}^2(\tilde F_1,\tilde F_2),
		\end{aligned}
		\end{eqnarray}
		and the on-site interaction constant
		\begin{eqnarray}
		\label{eq.Uo}
		\begin{aligned}
		U_e={U_0}-2V_e \ ,
		\end{aligned}
		\end{eqnarray}
		which differs from the constant $U_0$ of the static system by the contribution to the two nearest neighbour interaction processes.

	\subsubsection{Co-tunnelling}
	\label{sec.cotunneling}
Co-tunnelling processes with two particles tunnelling from one site $j$ to the same neighbour site $k$, can be described by
	\be
	\hat{H}_c=\sum_jT_cc_{k}^\dagger c_{k}^\dagger c_jc_j
	\ee
	with $k=j+1$ or $k=j-1$.
	The tunnelling rate can be expressed in Bessel functions as
		\begin{equation}
		\begin{aligned}
T_c=-\frac{j_0^2}{\omega^2}\sum_{n\neq 0}J_n(\tilde F_1,\tilde F_2)\left[\frac{U_d}{n}\frac{\partial}{\partial \tilde F_2}+\frac{U_0}{4n^2}\right]J_{-n}(\tilde F_1,\tilde F_2)\ .
\label{eq:cotunnelling}
	\end{aligned}
		\end{equation}
In contrast to the density assisted tunnelling discussed above, there is no preferred direction, and the rate for particles to tunnel to the left coincides with the rate for particles to tunnel to the right.

\subsubsection{Split-tunnelling}
\label{sec.split}
Another emerging process appearing in second order can be interpreted as split-tunnelling, where particles on the same site $j$ tunnel to two distinct adjacent sites, {\it i.e.} site $j-1$ and $j+1$ in the one-dimensional case.
This is captured by the Hamiltonian
	\be
\hat{H}_s=\sum_jT_sc_{j+1}^\dagger c_{j-1}^\dagger c_jc_j
\ee
with the split tunnelling rate $T_s$ that is determined by the interaction constant $V_e$ given in Eq.~\ref{eq.V} by the relation $T_s=-V_e/8$.
	
\subsubsection{Next-nearest-neighbour density assisted tunnelling}
\label{sec.NNNdat}

In second order ($1/\omega^2$), there is nearest-neighbour density assisted tunnelling
described by the Hamiltonian
\begin{eqnarray}
\hat{H}_a=\sum_jT_cc_j^{\dagger}\left(\hat{n}_j
-4\hat{n}_{j+1}+\hat{n}_{j+2}\right)c_{j+2}\ .
\end{eqnarray}
The amplitude for this process coincides indeed with the amplitude for co-tunnelling given above in Eq.~\eqref{eq:cotunnelling}

Similarly to the discussion
in Sec.~\ref{sec.nndat}, the amplitude $T_c$ should not be taken literally as a rate,
but rates for tunnelling processes depend on the occupation of all involved sites.
For the initial Fock state $\ket{\psi_i}=\ket{n_1\hdots n_N}$, one obtains the rates
\begin{eqnarray}
\begin{aligned}
T_cc_j^{\dagger}&\left(\hat{n}_j
-4\hat{n}_{j+1}+\hat{n}_{j+2}\right)c_{j+2}\ket{\psi_i} \\
&= T_c(n_j-4n_{j+1}+n_{j+2}-1) c_j^{\dagger} c_{j+2} \ket{\psi_i}\ ,
\end{aligned}
\end{eqnarray}
such that $T_c(n_j-4n_{j+1}+n_{j+2}-1)$ can be interpreted as a rate,
and, in contrast to the nearest-neighbour tunnelling discussed above, the rate for tunnelling to the right coincides with the rate to tunnel to the left.

\subsection{Control}
\label{sec.control}
We will be interested in the dynamics induced by nearest neighbour interactions and density assisted tunnelling in addition to the regular tunnelling and on-site interaction captured by the Bose-Hubbard Hamiltonian.

We focus on the resulting effective model
\begin{eqnarray}
\label{eq.H_eff}
\begin{aligned}
H_e&=\sum_{\langle jk\rangle}\hat{c}_j^{\dagger}J^{(e)}_{jk}\hat{c}_k+\sum_{\langle jk\rangle}T^{(e)}_{jk}\hat{c}_j^{\dagger}(\hat n_j-\hat n_k)\hat{c}_k\\
&+\frac{U_e}{2}\sum_j \hat{n}_j(\hat{n}_j-1)+\frac{V_e}{2}\sum_{\langle jk\rangle}\hat{n}_j\hat{n}_k,
\end{aligned}
\end{eqnarray}
which features the rich physics of different strongly correlated phases of matter, as we show in Sec. \ref{Sec.Phase}.

In order to arrive at this model, we need to ensure that split-tunnelling, co-tunnelling and next-nearest neighbour tunnelling vanish or become negligible.
Since the rate $T_s$ of split-tunnelling depends on the strength $V_e$ via the relation $T_s=-V_e/8$, it is not possible to find driving patterns resulting in vanishing $T_s$ but finite $V_e$.
The factor $1/8$, however, ensures that the rate of split tunnelling is an order of magnitude smaller than the nearest neighbour interaction strength, such that we neglect it in the following. 

In order to have a substantial impact of the nearest-neighbour interaction, it is important that it is not outweighed by the on-site interaction.
Consistent with the perturbative expansion, the ratio between the interaction strengths $U_e$ and $V_e$ following Eqs.~\eqref{eq.V} and \eqref{eq.Uo} can be approximated as
%
\be
\label{eq.ratio}
\frac{V_e}{U_e}\simeq\frac{4j_0^2}{\omega^2}\sum_{n\neq 0}\left[\frac{1}{2 n^2}-\frac{1}{n}\frac{U_d}{U_0}\frac{\partial}{\partial \tilde F_2}\right]J_{n}^2(\tilde F_1,\tilde F_2)\ .
\ee
Generically, the nearest-neighbour interaction will thus be substantially weaker than the on-site interaction due to the prefactor ${j_0^2}/{\omega^2}$.
It is however possible to chose the static interaction constant $U_0$ to be much smaller than the amplitude $\omega U_d$ of the oscillating interaction, and thus arrive at a sizeable ratio between $V_e$ and $U_e$. On the other hand, it is necessary to choose a non-zero $F_1$ to make sure the second term in Eq.~\ref{eq.ratio} remains finite, as discussed in more detail in Appendix~\ref{sec.special}.

A finite driving amplitude $F_2$ is not necessary for the realisation of nearest neighbour interactions,
but it is required for the realisation of an extended Bose Hubbard model with the parity symmetry breaking with $h_{(1,1)}^{R}\neq h_{(1,1)}^{L}$.
In  order to arrive at strongly correlated ground states, the ratio between the nearest neighbour interaction and the nearest neighbour tunnelling rates needs to be sufficiently large.
Following Eqs.~\ref{equ.channel} and \ref{eq.V} the ratio of these two rates can be approximated as
		\begin{eqnarray}
\label{eq.Vhratio}
\begin{aligned}
\frac{V_e}{h_{(0,1)}^{L}}\approx\frac{4j_0U_d}{\omega^2 J_0(\tilde F_1,\tilde F_2)} \sum_{n\neq 0} \frac{1}{n}\frac{\partial}{\partial \tilde F_2}J_{n}^2(\tilde F_1,\tilde F_2)\ ,
\end{aligned}
\end{eqnarray}
if $U_0$ is much smaller than $U_d$.
The factor $j_0U_d/\omega^2$ is generically small, but the terms involving Bessel functions can be tuned within a wide range, and especially a substantial ratio can be realised with a small value of $J_0(\tilde F_1,\tilde F_2)$.

\begin{figure}[h]
	 			\centering
 	\includegraphics[width=0.5\textwidth]{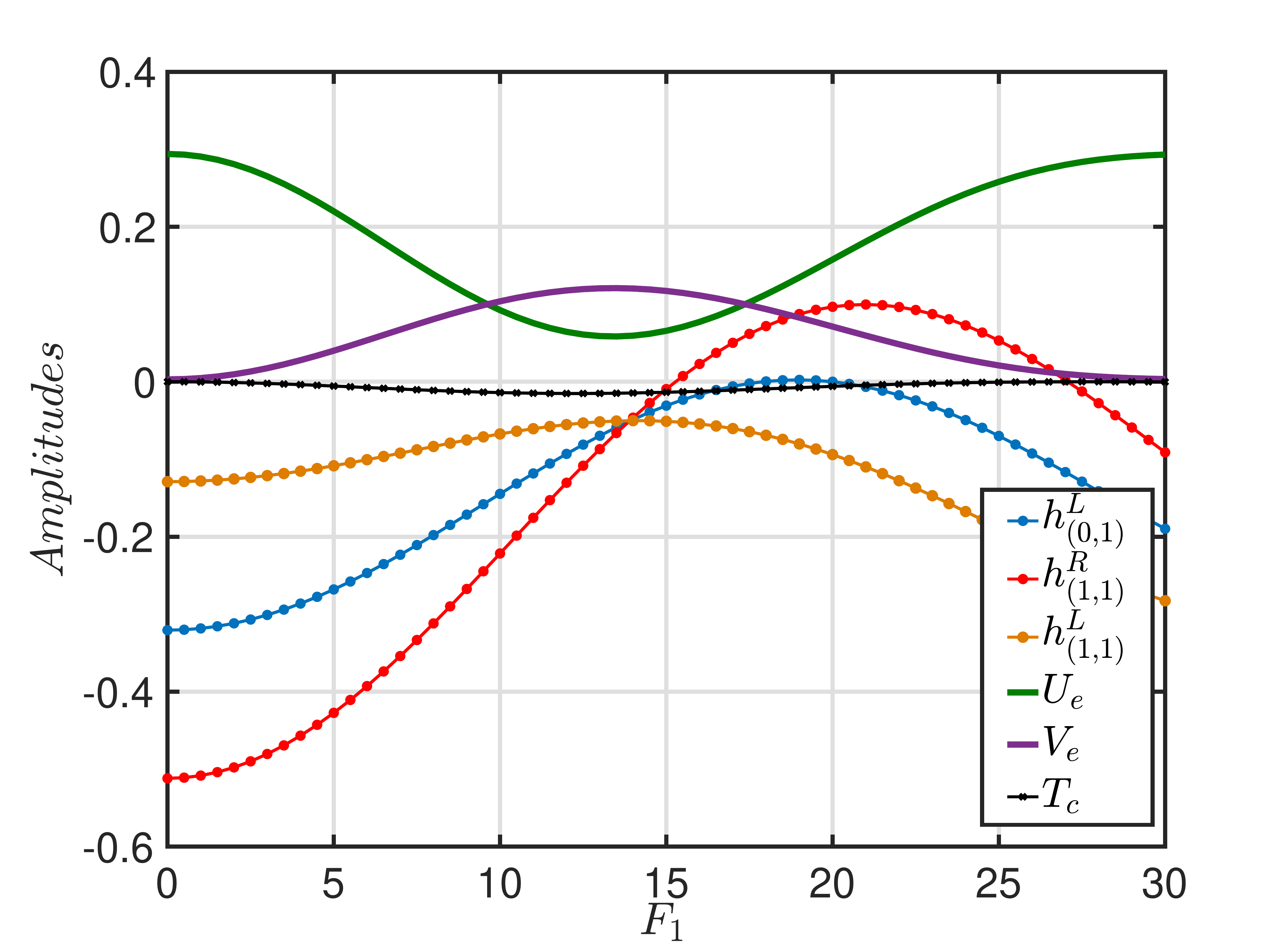}
	 	\caption{: Amplitudes of selective tunnelling channels and interactions for one varying driving field $F_1$,  with fixed $j_0=1,\omega = 6j_0,U_d=-4j_0,U_0=0.3j_0,F_2=22j_0$. Induced repulsive interactions can be tuned to be notable from 10 to 17. The occupation number dependency for different tunnelling channels can be clearly observed, and $h_{(1,1)}^R$ can be suppressed to zero around $F_1=15.5$ while others remain finite. $T_c$ is negligible.
	 		}
 			 	\label{fig.amp}
	 \end{figure}

With these general observations in mind, we can now discuss the amplitudes for the effective processes and their dependence on the tuneable parameters $U_d$, $F_1$ and $F_2$ defined above in Eqs.~\eqref{eq:parameterU} and \eqref{eq:parameterF}.
Fig.~\ref{fig.amp} depicts the amplitudes (in terms of the bare tunnelling rate $j_0=1$) for several effective processes for driving frequency $\omega=6j_0$, 
a static on-site interaction $U_0=0.3j_0$ and an AC on-site interaction amplitude $U_d= -4j_0$.
The amplitude $F_2$ adopts the value $F_2=22j_0$ and $F_1$ ranges between $0$ and $30j_0$.
The rate $V_e$ of effective nearest-neighbour interaction (purple) is negligible for small values of $F_1$ as discussed above in Sec.~\ref{sec.control}.
In the parameter window $5j_0\lesssim F_1\lesssim 22j_0$, however, the nearest-neighbour interaction is by no means negligible as compared to the other processes;
in particular around the sweet-spot $F_1\simeq 14j_0$ the rate of nearest-neighbour interaction is approximately twice as large as the prefactor of on-site interaction.

In addition, tunnelling processes can be suppressed with suitable choices of $F_1$.
In particular for $F_1\simeq15.5j_0$, the tunnelling process with rate $h_{(1,1)}^R$ is forbidden, but the remaining channels still exist with the ratio $|V_e/h_{(1,0)}^{L/R}|\approx 5$. Another interesting point is around $F_1\approx 14$ such that $\frac{\partial }{\partial \tilde F_2}J_0(\tilde F_1,\tilde F_2)\approx0$, where three density-dependent tunnelling rates coincide at $-j_0 J_0(\tilde F_1,\tilde F_2)$ and  the parity symmetry remains preserved. 
The co-tunnelling rate $T_c$, depicted in black in Fig.~\ref{fig.amp}, is negligible for all values of $F_1$.
 
As the main result of our paper we have thus shown that the combination of shaking and modulated on-site interactions
significantly extends the achievable effective Hamiltonian,
and the individual terms can be enhanced or suppressed simply by tuning the driving parameters.
After neglecting suppressible undesired processes, the effective Hamiltonian up to the order of $\omega^{-2}$ can be written as Eq.~\ref{eq.H_eff}.
\section{Quantum phases}
\label{Sec.Phase}

Having discussed the possibility to tune the rates of various processes, we can finally give an idea of the type of physics that can be explored with the processes described above.
To this end, we consider the Hamiltonian
\begin{eqnarray}
\begin{aligned}
\hat{H} = &-j_0\sum_{j}\hat{c}_j^{\dagger}\hat{c}_{j+1}+ T_e \hat{c}_j^{\dagger}(\hat{n}_j-\hat{n}_{j+1})\hat{c}_{j+1} +h.c.\\
&+\frac{U_e}{2}\sum_j \hat{n}_j(\hat{n}_j-1)+V_e\sum_j \hat{n}_j\hat{n}_{j+1}\ ,
\end{aligned}
\label{equ.eff}
\end{eqnarray}
comprised of a bare Bose-Hubbard Hamiltonian, and additional density assisted tunnelling and nearest-neighbour interaction. 

In order to understand the ground state properties of this Hamiltonian,
it is helpful to consider the dynamics induced by $\hat{H}$ with the Mott insulator state $|\Psi\rangle = \bigotimes_{i=1}^{N} c_i^{\dagger}|0\rangle $ as initial state.
The transition amplitudes, defined in Eq.~\ref{equ.channel} -- expressed in terms of $j_0$ and $T_e$ -- read
\begin{eqnarray}
	\begin{aligned}
	h_{(0,1)}^L&=h_{(1,0)}^R = -j_0,\\
	h_{(1,1)}^L &= -j_0+T_e,\\
	h_{(1,1)}^R &= -j_0 - T_e.
	\end{aligned}
\end{eqnarray}
In particular for $T_e\simeq j_0$, the tunneling rates $h_{(1,1)}^L$ vanishes whereas the amplitude $h_{(1,1)}^R$ becomes particularly large.
The system will thus evolve from the Mott insulator state $|\Psi\rangle$ towards the state
\begin{equation}
|\Psi\rangle+2iT_et\sum_j\hat{c}_{j+1}^{\dagger}\hat{c}_{j}|\Psi\rangle+O(t^2)\ .
\end{equation}
The resulting imbalanced doublon-holon defect pair, with a single doubly occupied site and adjacent un-occupied site on its left, leads to violation of parity symmetry.
The existence of defect pairs reduces the nearest-neighbour interaction in the system Hamiltonian (Eq.~\eqref{equ.eff}), but it increases the on-site interaction energy.
One might thus expect that the system ground state shows signatures of a density wave for a suitable balance of the several processes in the system.

 We introduce the average number of two types of defect pairs in 02 or 20 order in the ground state as $N_{02/20}$ to assess it more quantitatively.
The difference
\be
\Delta N=N_{02}-N_{20}
\label{eq:defectpairdifference}
\ee
for defect pairs indicates a preferred direction of tunnelling.
\begin{figure}[h]
	\centering
	\includegraphics[width=0.5\textwidth]{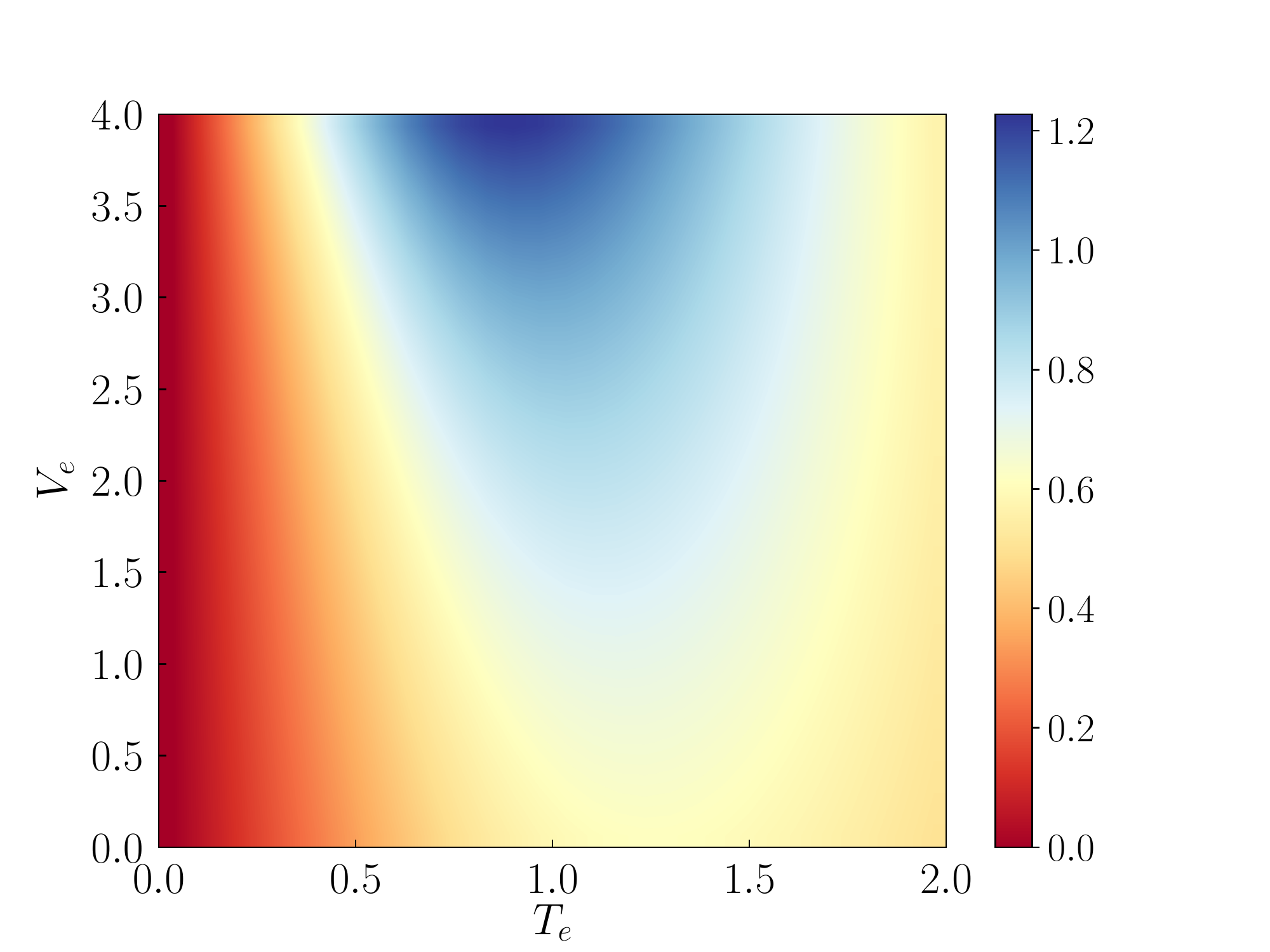}
	\caption{
		Defect pair difference (Eq.~\eqref{eq:defectpairdifference}) as a function of density assisted tunnelling amplitude $T_e$ and nearest neighbour interaction strength $V_e$ for a system with $8$ particles on $8$ lattice sites and a fixed on-site interaction $U_e=7j_0$.
		For sufficiently weak nearest neighbour interaction ($V_e\lesssim 3$), there is a broad interval $0.5j_0\lesssim T_e\lesssim 1.5j_0$ around of pronounced enhancement of defect pairs resulting from tunnelling to the right.}	
	\label{fig.P_diff}
\end{figure}
The creation of a pronounced imbalance between the two different types of defect pairs around the the tunnelling amplitude $T_e\simeq j_0$ can be seen in Fig.~\ref{fig.P_diff}, where $\Delta N$ is depicted based on numerically exact diagonalisation of the extended Bose Hubbard Hamiltonian (Eq.~\eqref{equ.eff}) for $8$ particles on $8$ lattice sites with periodic boundary conditions to reduce finite size effect.
Given the strong on-site interaction with $U_e=7j_0$, the system is in a Mott insulating state for sufficiently low values of $T_e$ and $V_e$.
The blue area shows the expected preference of the system for defects resulting for tunnelling to the right, and the number of defect pairs is expected to grow with increasing nearest-neighbour interactions.

In order to analyse the phase transition from the Mott insulator to a density wave more accurately,
we focus on the order parameter~\cite{EBHM}
	\begin{equation}
	R_{DW}=\lim\limits_{r\to\infty}(-1)^{r}\langle\delta n_j\delta n_{j+r}\rangle\ ,
	\label{eq:orderparameter}
	\end{equation}
where $\delta n_j = n_j-\langle n \rangle$ characterizes the deviation of the atom number on site $j$ from its average.
Due to periodic boundary conditions, the system has translation symmetry; the order parameter $R_{DW}$ is independent of $j$, and the prefactor $(-1)^{r}$ ensures that $R_{DW}$ is non-negative. For a density wave, the order parameter remains finite for large distances, but it vanishes for the Mott insulator.
In practice, we will chose $r=5$ which is the largest possible length within  $10$ sites.
	\begin{figure}
		{%
			\includegraphics[width=0.45\textwidth]{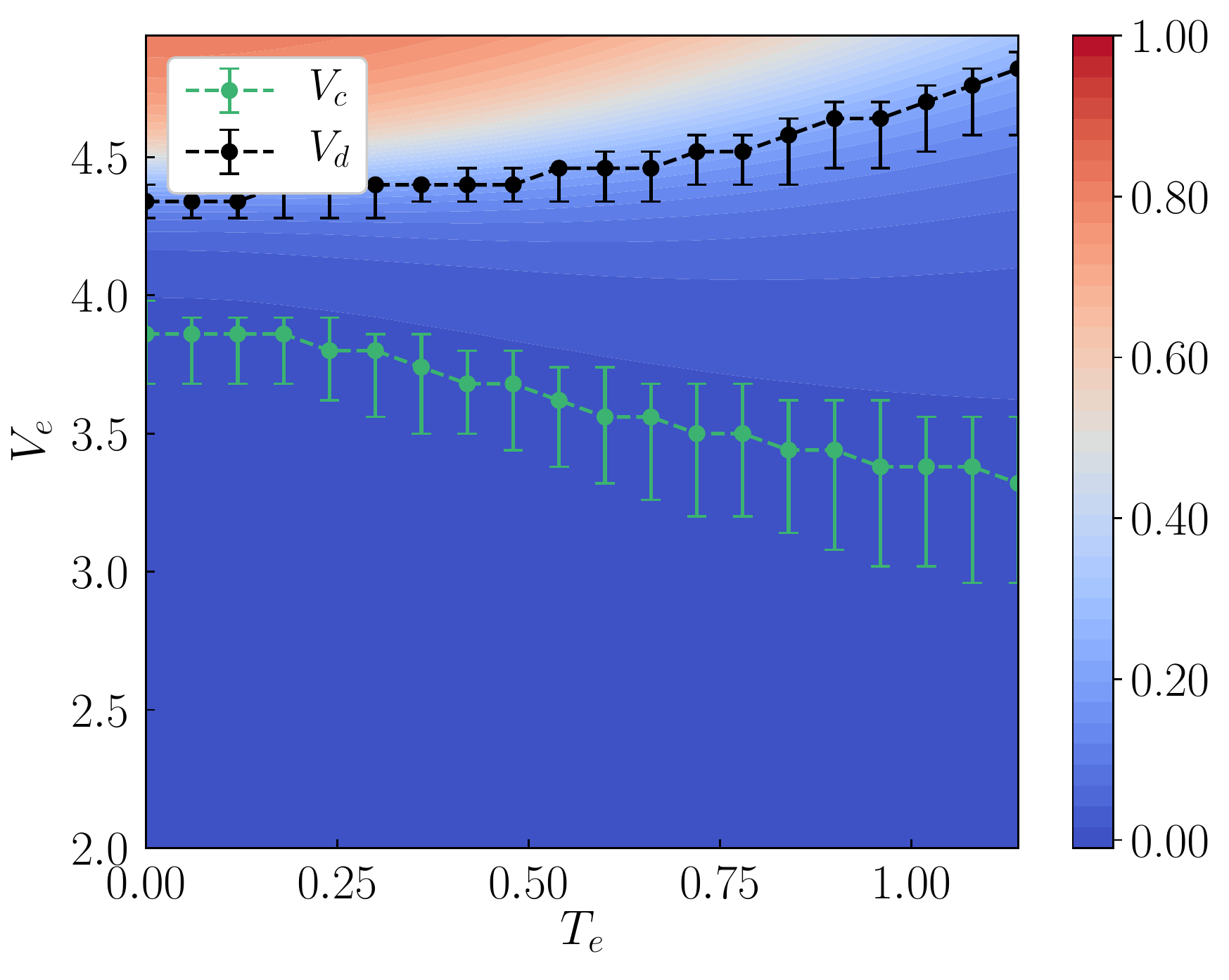}
		}\vfill
		{%
			\includegraphics[width=0.45\textwidth]{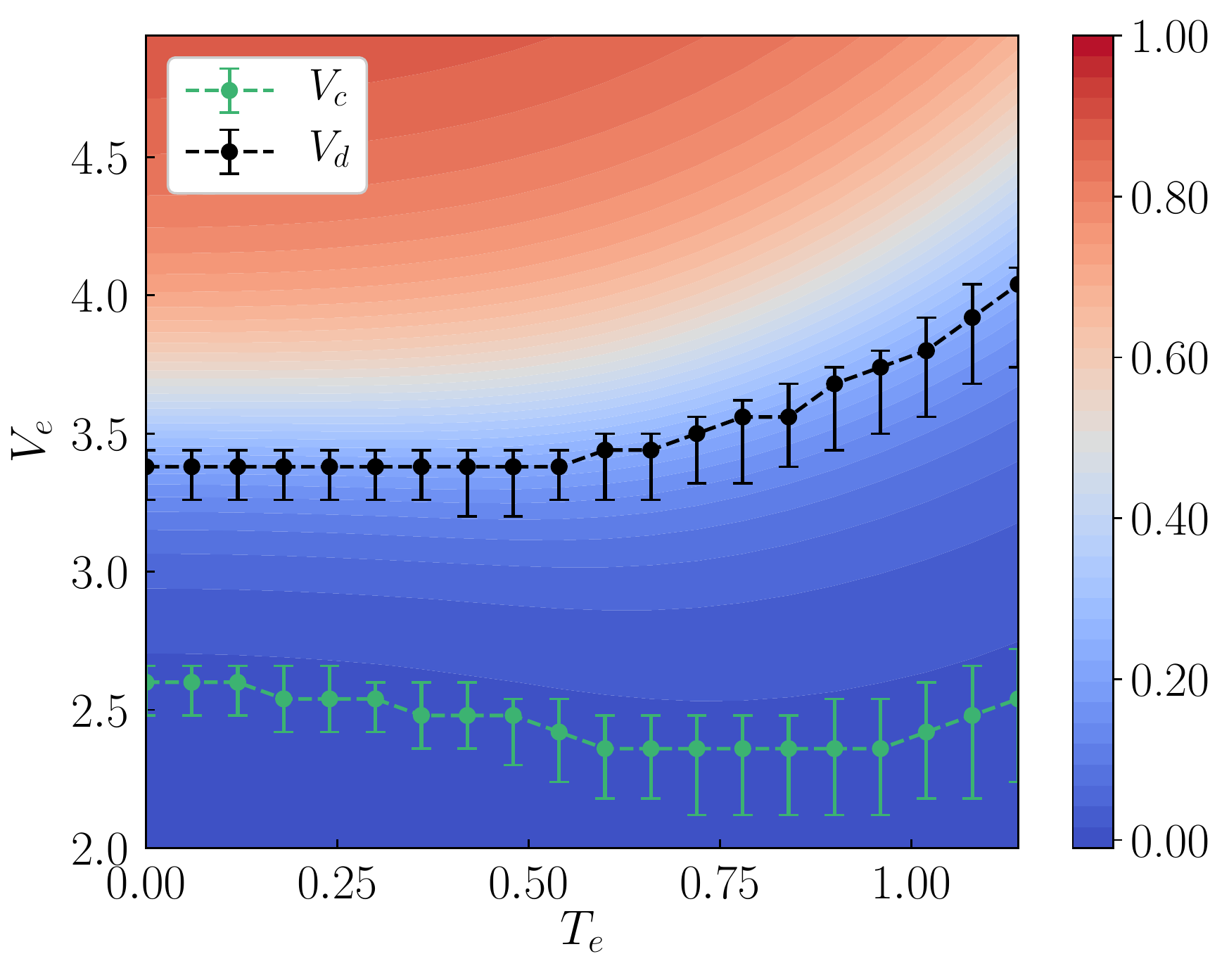}
		}
		\caption{Density wave order parameter $R_{DW}$ (Eq.~\eqref{eq:orderparameter}
of the ground state of the extended Bose Hubbard Hamiltonian (Eq.~\eqref{equ.eff}) for $10$ particles on $10$ lattice sites for an on-site interaction of strength $U_e=10j_0$ (top) and $U_e=5j_0$ (bottom).
Red colors indicate a clear density wave and blue colors indicate similarity to a Mott insulating state.
The depicted phase boundaries correspond to $R_{DW}=0.25\pm0.05$ (black) and $R_{DW}=0.01\pm0.005$
(green).
The domain between those boundary lines corresponds to a smooth change of system properties due to the finite system size.
As one can see, strong nearest neighbour interactions favor the formation of a density wave, whereas strong tunnelling and strong on-site interactions (comparison of top and bottom) inhibit its formation.}

\label{fig.DW}
\end{figure}

This order parameter is plotted as function of density assisted tunnelling rate $T_e$ and nearest neighbour interaction constant $V_e$ in Fig.~\ref{fig.DW}.
Given the finite system size, there is not a sharp transition from Mott insulator (blue) to a density wave (red),
but the threshold values $R_{DW}=0.25\pm0.05$ and $R_{DW}=0.01\pm0.005$ depicted as $V_d$ (black) and $V_c$ (green) provide an estimate for the transition.
The establishment of a density wave is clearly supported by strong nearest neighbour interactions, whereas density-assisted tunnelling tends to destroy the density wave like behaviour.
The top figure corresponds to an on-site interaction of strength $U_e=10j_0$ whereas the bottom figure corresponds to the weaker on-site interaction with $U_e=5j_0$.
The comparison of both figures confirms that also strong on-site interactions tend to break density wave like behaviour as expected.

\section{Summary}

Many experiments over the last decade highlighted the technological advances towards the control that is required for the realization of a quantum simulator.
After those developments, it is now possible to not only control single-particle tunnelling processes, but to also introduce artificial interactions.
This control over interactions pushes quantum simulations from proof-of-concept experiments on single-body physics to the exploration of many-body physics which would be strongly limited in terms of classical computational approaches.

The possibility to modulate both on-site energies and tunnelling processes as considered here offers quite some potential for the realisation of effective processes beyond single-body physics.
Those processes help to realise correlated many-body phases like density waves as discussed here,
but the range of accessibly many-body ground states is substantially wider than we would be able to explore here, e.g. the Haldane Bose insulator with unconventional string order~\cite{HI}. 
In addition,  in two-dimensional lattices, where the interplay of different tunnelling processes can lead to the formation of states with topological order, there is an abundance of conceivable reactions of many-body system to different interaction effects. 
Exploring those in actual quantum simulations would allow us to actually exceed the limitation of our classical computers for questions of fundamental science.

In this article we focus on the engineered nearest neighbour interactions, a non-trivial extension of this work will be realizing notable longer ranged interactions. As seen from Eq.~\ref{eq.longrange}, the induced interactions are not limited to nearest neighbouring sites, as long as tunnellings beyond nearest neighbours are permitted, e.g. for quantum gas systems in shallow lattices. Consequently more exotic strongly correlated phases are probably within reach\cite{NS}.

As we demonstrated here, the driving schemes to realize such quantum simulations do not require sophisticated shaped pulses, but simple bi-chromatic driving already gives us access to a broad range of physical processes.

	\section{Acknowledgement}
	
	We are indebted to Frederic Sauvage for stimulating discussions.
	
	\appendix
	\section{Effective Hamiltonian }
	\label{Sec.effectiveH}
    Physical origins and derivations of effective processes illustrated in Sec.~\ref{Sec.process} will be discussed here.
	In the high-frequency regime in which the driving frequency $\omega$ exceeds all other relevant scales in the system Hamiltonian, 
	 the time evolution can be approximately captured by the time-independent effective Hamiltonian which can be obtained perturbatively in powers of $1/\omega$.  For the present purposes, terms up to $1/\omega^2$ will be sufficient.
	 In first order ($\propto1/\omega$), two-step tunnelling results from a particle tunnelling from site $k$ to site $m$ and moving to site $j$ immediately after, as reflected by the commutation relation 
	\begin{eqnarray}
	[c_j^{\dagger}c_m,c_m^{\dagger}c_k]=c_j^{\dagger}c_k\ .\nonumber
	\end{eqnarray}
The process in which the particle tunnels back to site $k$ ({\it i.e.} $j=k$) results in a modification of the on-site energy.
	
	In addition, two-step density assisted tunnelling($1/\omega^2$) appear resulting from a particle tunnelling from site $k$ to site $m$, interacting with the particles at site $m$ and keep moving to site $j$,
	as reflected by the commutation relation
	\begin{equation}
	\left[c_j^\dagger c_m,\left[\hat n_m(\hat n_m-1),c_m^\dagger c_k\right]\right]=4c_j^{\dagger}n_mc_k .\nonumber
	\end{equation} In fact, besides interacting with particle at site $m$, the interaction can happen at site $j,k$ as well, therefore the occupation number dependence also takes additional form of $c_j^{\dagger}n_kc_k,c_j^{\dagger}n_jc_k$.   Finally split-tunnelling results from two particle first interacting with other particles at the same site $m$, then tunnelling to distinct nearest neighbour sites $j$ and $k$ respectively as 
	\begin{equation}
	\left[c_k^\dagger c_m,\left[\hat n_m(\hat n_m-1),c_j^\dagger c_m\right]\right]=-2c_j^{\dagger}c_k^{\dagger}c_mc_m .\nonumber
	\end{equation}
	The co-tunnellings is almost the same as split-tunnelling apart from that two particles  ending at same site as $c_j^{\dagger}c_j^{\dagger}c_mc_m $. Finally the engineered interactions have been discussed  in Sec.~\ref{Sec.Eff}. In order to get the amplitudes for each process, in practice we use Magnus expansion to compute the lowest three orders effective Hamiltonians, and within each of them we neglect all terms with higher order than $1/\omega^2$.
	\paragraph{Lowest order effective Hamiltonian}
The lowest effective Hamiltonian is the temporal average of the time-dependent Hamiltonian (Eq.~\eqref{equ.H}), and reads
	\begin{eqnarray}
	\hat{H}_0 =\sum_{\langle jk \rangle}\hat{c}_j^{\dagger}\hat{A}_{jk}^0\hat{c}_k+\frac{U_0}{2}\sum_{j}\hat{n}_j(\hat{n}_j-1),
	\end{eqnarray}
	where the density assisted tunnelling can be expressed as
	\begin{equation}
	\hat{A}_{jk}^{0}=\frac{1}{T}\int_0^T  J_{jk}e^{i\chi_{jk}(t)}e^{i\Gamma(t)(\hat{n}_j-\hat{n}_k)}dt.
	\end{equation}
In the one dimensional case with the drivings defined in Eq.~\ref{eq:parameterU}~\ref{eq:parameterF},
 it reduces to
	\begin{equation}
\hat{A}_{jk}^{0}=-j_0 J_0((j-k)\tilde{F}_1,(j-k)\tilde{F}_2+\tilde{U}_d(\hat{n}_j-\hat{n}_{k})),
\end{equation}
and the dependence on density difference in terms of $1/\omega$ can be derived from Eq.~\ref{eq.A_expand}.

	\paragraph{First order effective Hamiltonian}
	The construction of the first order effective Hamiltonian is eased by expressing the Hamiltonian $H(t)$ in terms of its Fourier components $\hat{H}_{n}=\sum_{\langle jk\rangle}\hat{c}_j^{\dagger}\hat{A}_{jk}^n\hat{c}_k,$ with 
	 	\begin{equation}
	 \hat{A}_{jk}^{n}=\frac{1}{T}\int_0^T  J_{jk}e^{i\chi_{jk}(t)}e^{i\Gamma(t)(\hat{n}_j-\hat{n}_k)}e^{-in\omega t}dt\ ,
	 \end{equation}
	 for $n\neq 0$.
Up to first order, the operators $\hat{A}_{jk}^{n}$ can be approximated as $\hat{A}_{jk}^{n}\approx g_{jk}^n+(\hat{n}_j-\hat{n}_k)t_{jk}^n$
with the Fourier components 
	 		\begin{eqnarray}
	 \begin{aligned}
	 g_{jk}^n&=\frac{1}{T}\int_0^T J_{jk}e^{i\chi_{{jk}}(t)}e^{-in\omega t}dt\\
	 t_{jk}^n&=\frac{i}{T}\int_0^TJ_{jk}e^{i\chi_{{jk}}(t)}\Gamma(t)e^{-in\omega t}dt\ .
	 \end{aligned}
	 \end{eqnarray}
	 is sufficient.
	 For the  one dimensional case, we have
	\begin{equation}
	\label{eq.gt}
{g}^n_{jk} =
-j_0J_n(\tilde{F}_1(j-k),\tilde{F}_2(j-k)), t_{jk}^n = \frac{\tilde{U}_d}{j-k}\frac{\partial g_{jk}^n}{\partial \tilde{F}_2}
\end{equation} 
where $k=j\pm 1$ for nearest neighbour tunnelling.
The first order effective Hamiltonian including terms up to second order thus reads
	\begin{eqnarray}
	\label{eq.H_eff_general}
	\begin{aligned}
	&\hat{H}_{eff}^1 = \frac{1}{\omega}\sum_{n=1}^{\infty}\frac{1}{n}[\hat{H}_n,\hat{H}_{-n}]\\
	&\approx\sum_{\langle\langle jk\rangle\rangle,j=k}c_j^{\dagger}B_{jk}c_k+\big[\sum_{jmk}T_{jmk}c_j^{\dagger}c_{k}^{\dagger}c_mc_m+h.c.\big]\\
	&+\sum_{\langle\langle jk\rangle\rangle}c_j^{\dagger} \hat{S}_{jk}c_k +\sum_{\langle  jk\rangle}\frac{U_{jk}^{N}}{2}\hat{n}_j\hat{n}_k-\sum_j \frac{U_j^{O}}{2} \hat{n}_j(\hat{n}_j-1),
	\end{aligned}\\
	\end{eqnarray}
with the coefficients
	\begin{eqnarray}
	\label{eq.coe}
	\begin{aligned}
	&B_{jk} = \sum_{n\neq 0}\frac{1}{n\omega}\sum_mg_{jm}^ng_{mk}^{-n}\ ,\\ 
	&T_{jmk}=\sum_{n\neq 0}\frac{1}{n\omega}(g_{jm}^{-n}t_{km}^n)\ ,\\
	&U_{jk}^N=-4(S^1_{jkj}+S^2_{jkj}), U_j^O = \sum_kU_{jk}^N/2\ ,\\
	&\hat{S}_{jk}=\sum_{m}\big[S_{jmk}^{1} \hat{n}_j
	-2(S_{jmk}^{1}+S_{jmk}^{2})\hat{n}_m+S_{jmk}^{2n}\hat{n}_k\big]\ ,\\
	&S_{jmk}^{1}= \sum_{n\neq 0}\frac{1}{n\omega}t_{jm}^ng_{mk}^{-n},\ S_{jmk}^{2}=\sum_{n\neq 0}\frac{1}{n\omega}g_{jm}^{-n}t_{mk}^{n}.
	\end{aligned}
	\end{eqnarray}
	
	The first term of Eq.~\ref{eq.H_eff_general}, $\sum_{\langle\langle jk\rangle\rangle,j=k}c_j^{\dagger}B_{jk}c_k$ indicates the two-step tunnelling and the modification to  on-site energies with rates $B_{jk}$. As proven in the Appendix.A in ~\cite{SD}, both of them vanish regardless of the drivings for the one dimensional case. The second expression of Eq.~\ref{eq.H_eff_general} with rate $T_{jmk}$ illustrates the  co-tunnelling and split-tunnelling processes. The operator $\sum_{\langle\langle jk\rangle\rangle}c_j^{\dagger} \hat{S}_{jk}c_k$ captures the two-step density-assisted tunnelling. The last two give us the induced nearest neighbour interactions and the modification to the on-site interaction. 
	\paragraph{Second order effective Hamiltonian}
	The second order processes ($1/\omega^2$) also exist in the second order effective Hamiltonian $H_{eff}^2$,which can be expressed \cite{HF}  as
	\begin{equation}
	\label{eq.H_eff2}
	\hat{H}_{eff}^2 = \sum_{n\neq 0}\Big( \frac{[\hat{H}_{-n},[\hat{H}_0,\hat{H}_n]]}{2n^2\omega^2}
	+\sum_{n'\neq 0,n}\frac{[\hat{H}_{-n'},[\hat{H}_{n'-n},\hat{H}_n]]}{3nn'\omega^2}\Big).
	\end{equation}
	 
	Since $\omega^{2}$ appears in the dominator,
	all contributions with terms $\hat{H}_{n}\propto 1/\omega$ are negligible
	and we can make the following approximation 
	\begin{eqnarray}
	\begin{aligned}
	&\hat{H}_0  \approx\sum_{\langle jk\rangle}\hat{c}_j^{\dagger}g_{jk}^0\hat{c}_k+\frac{U_0}{2}\sum_{j}\hat{n}_j(\hat{n}_j-1),\\
	&\hat{H}_{n\neq 0}\approx\sum_{\langle jk\rangle}\hat{c}_j^{\dagger}g^n_{jk}\hat{c}_k.
	\end{aligned}
	\end{eqnarray}
Such an approximation leads to the crucial simplification to compute $\hat{H}_{eff}^2$ by noticing that the commutator appearing in Eq.~\ref{eq.H_eff2}
\begin{eqnarray}
[\sum_{\langle jk\rangle}\hat{c}_j^{\dagger}g^p_{jk}\hat{c}_k,\sum_{\langle jk\rangle}\hat{c}_j^{\dagger}g^q_{jk}\hat{c}_k] = \sum_{\langle \langle jk\rangle\rangle,j=k}c_i^{\dagger}B_{jk}^{pq}c_k,
\end{eqnarray}
with $B_{jk}^{pq} =  \sum_l g_{jl}^p g_{lk}^q-g_{jl}^qg_{lk}^p$, vanishes in the one dimensional case for arbitrary $p,q$. For instance, when $k=j+2$, the tunnelling rate reduces to $$B_{j,j+2}^{pq} = g_{j,j+1}^p g_{j+1,j+2}^q-g_{j,j+1}^qg_{j+1,j+2}^p,$$ which turns to be zero according to Eq.~\ref{eq.gt}.
Therefore the non-vanishing contribution to $\hat{H}^2_{eff}$ is merely from the presence of the constant on-site interaction $U_0$ as
\begin{equation}
\hat{H}_{eff}^2 = \sum_{n\neq 0} \frac{[\hat{H}_{-n},[\hat{H}_{int},\hat{H}_n]]}{2n^2\omega^2},
\end{equation}
with the interacting Hamiltonian $\hat{H}_{int}=\frac{U_0}{2}\sum_{j}\hat{n}_j(\hat{n}_j-1)$. No new process will be introduced, but the amplitudes given in Eq.~\ref{eq.coe} will be modified as
\begin{eqnarray}
\label{eq.secondco}
\begin{aligned}
&\tilde{T}_{jmk}=T_{jmk}-\sum_{n\neq 0}\frac{U_0}{4n^2\omega^2}(g_{jm}^{-n}g_{km}^n),\\
&\tilde{S}_{jk}=\sum_{m}[\tilde{S}_{jmk}^{1} \hat{n}_j
-2(\tilde{S}_{jmk}^{1}+\tilde{S}_{jmk}^{2})\hat{n}_m+\tilde{S}_{jmk}^{2}\hat{n}_k]
\end{aligned}
\end{eqnarray}
with 
\begin{eqnarray}
\begin{aligned}
\tilde{S}_{jmk}^{1}={S}_{jmk}^{1}-\sum_{n\neq 0}\frac{U_0}{4n^2\omega^2}(g_{jm}^{n}g_{mk}^{-n})\\
\tilde{S}_{jmk}^{2}={S}_{jmk}^{2}-\sum_{n\neq 0}\frac{U_0}{4n^2\omega^2}(g_{jm}^{n}g_{mk}^{-n})
\end{aligned}
\end{eqnarray}
where $S_{jmk}^{1/2},T_{jmk}$ are defined in Eq.~\ref{eq.coe}.
For the one dimensional case, those reduce to the co-tunnelling rate
\begin{equation}
\begin{aligned}
T_c&=\sum_{n\neq 0}\frac{1}{n\omega}g_{j+1,j}^{-n}t_{j+1,j}^n-\sum_{n\neq 0}\frac{U_0}{4n^2\omega^2}g_{j+1,j}^{-n}g_{j+1,j}^n,
\end{aligned}
\end{equation}
the split tunnelling rate
\begin{equation}
\begin{aligned}
T_s&=\sum_{n\neq 0}\frac{1}{n\omega}g_{j+1,j}^{-n}t_{j-1,j}^n-\sum_{n\neq 0}\frac{U_0}{4n^2\omega^2}g_{j+1,j}^{-n}g_{j-1,j}^n,
\end{aligned}
\end{equation}
and the interaction constant.
$V_e = -8T_s$
According to Eq.~\ref{eq.gt}, one can express in terms of Bessel functions and retrieve expression given in Sec.~\ref{Sec.Eff}.

  \section{Special case with zero $F_1$}
  \label{sec.special}
 
In order to achieve a notable ratio of $V_e/U_e$ as considered in Sec.~\ref{sec.control}, non-vanishing $F_1$ is demanded, otherwise the contribution proportional to $U_d$ to the nearest neighbour interactions(Eq.~\ref{eq.V}) vanishes. According to Eq.~\ref{eq.V}, the first term involving $U_d$ reads $$V_d=-\frac{4j_0^2U_d}{\omega^2}\sum_{n \neq 0}\frac{\partial}{\partial \tilde{F}_2}\frac{1}{n}J_n^2(\tilde{F}_1,\tilde{F}_2).$$
All contributions with odd $n$ vanishes since the Bessel function $J_n(0,\tilde{F}_2)$ is zero for all values of $\tilde{F}_2$ according to the symmetry relation of the two dimensional Bessel function~\cite{Bessel}.   $V_d$ reduces to 
 \begin{eqnarray}
 \begin{aligned}
V_d=\frac{\partial}{\partial \tilde{F}_2}\sum_{n =2k}^{+\infty}\frac{1}{2k}(J_{2k}^2(0,\tilde{F}_2)-J_{-2k}^2(0,\tilde{F}_2))
 \end{aligned}
 \end{eqnarray}
 where the two dimensional Bessel function can be rewritten as a one dimensional Bessel function as $J_{-2k}(0,\tilde{F}_2) = J_k(\tilde{F}_2)$. Finally the property $J_{-k}(\tilde{F}_2)=(-1)^kJ_k(\tilde{F}_2)$\cite{Bessel} leads to the expression
 \begin{eqnarray}
 \begin{aligned}
V_e=\frac{\partial}{\partial \tilde{F}_2}\sum_{n =2k}^{+\infty}\frac{1-(-1)^{2k}}{2k}J_{k}^2(\tilde{F}_2),
 \end{aligned}
 \end{eqnarray}
where the numerator $1-(-1)^{2k}$ cancels. Therefore a finite $\tilde{F}_1$ is crucial to generate $V_d$ and furthermore the notable nearest neighbour interactions.

	\bibliography{reference}
	\bibliographystyle{unsrt}
\end{document}